\documentclass[a4paper,12pt]{article}
\usepackage{graphicx,amssymb}
\renewcommand{\figurename}{Fig.}

\title{Isovector energy-weighted sums for hot nuclei in presence of
relaxation processes}

\author{V.M. Kolomietz\thanks{E-mail: vkolom@kinr.kiev.ua},\
S.V. Lukyanov\thanks{E-mail: lukyanov@kinr.kiev.ua}, 
O.I. Davidovskaya\thanks{E-mail: oid@kinr.kiev.ua} \\
{\it Institute for Nuclear Research, 03680 Kyiv, Ukraine}}

\begin{document}
\maketitle

\begin{abstract}
We investigate the collective response function and the energy-weighted sums 
$m_k$ for isovector mode in hot nuclei. The approach is based on the collisional 
kinetic theory and takes into consideration the temperature and the relaxation 
effects. Taking into account a connection between the isovector sound mode and 
the corresponding surface vibrations we have established the $A$-dependence of 
the enhancement factor for the isovector sum rule in a good agreement with 
experimental data. We have shown that the enhancement factor for the "model 
independent" sum $m_1$ is only slightly sensitive to the temperature change.
\vspace{1pc}


\end{abstract}

\section{Introduction}

Many features of nuclei are sensitive to nuclear heating. The nuclear
heating influences strongly the particle distribution near the Fermi surface
and reduces the Fermi-surface distortion effects on the nuclear collective
dynamics \cite{shko05}. Moreover, the heating of the nucleus provides the transition
from the rare- to frequent interparticle collision regime. One can expect
that the zero-sound excitation modes which exist in cold nuclei will be
transformed to the first-sound ones in hot nuclei. Knowledge of the nuclear
collective dynamics in hot nuclei allows one to understand a number of
interesting phenomena, e.g., the temperature dependence of the basic
characteristics of isovector giant dipole resonance (IVGDR).

A good first orientation in a description of the collective dynamics in hot
nuclei is given by a study of the response function and the energy weighted
sums in a nuclear matter within the kinetic theory \cite{lipi}. The nuclear matter
exhibits the properties of a Fermi liquid \cite{abkh59} and its description within the
kinetic theory requires the use of an effective nucleon-nucleon interaction
like Skyrme forces, Landau interaction, etc. We apply the Landau's kinetic
theory to the evaluation of the response function and the energy-weighted
sums in a two-component nuclear Fermi liquid. Both the temperature and the
relaxation phenomena are taken into account.

\section{Response function within the kinetic theory}

To derive the energy-weighted sums for the isovector excitations, we will
consider the density-density response of two-component nuclear matter to the
following external field
\begin{equation}\label{eq1}
U_{ext}(t)=\lambda_0 e^{-i(\omega + i0)t}\widehat{q} 
+ \lambda_0^*  e^{i(\omega - i0)t}\widehat{q}^{\ \ast},
\end{equation}
where $\lambda_0$ is the small amplitude, $\widehat{q}$ is the one-body operator 
$\widehat{q}=\sum\limits_{j=1}^{A}\widehat{q}(\vec{r}_j, \xi_j)
=\sum\limits_{j=1}^A \xi_j e^{-i\vec{q}\cdot\vec{r}_j}$ 
and $\xi_j$ is the isotopic index. The response density-density function 
$\chi (\omega)$ is given by \cite{lali}
\begin{equation}\label{chi1}
\chi(\omega)=\frac{\left\langle e^{-i\vec{q}\cdot\vec{r}}\right\rangle}
{\lambda_{0}e^{-i\omega t}}=\frac{1}{\lambda_{0}e^{-i\omega t}}\int d\vec{r}\
e^{-i\vec{q}\cdot\vec{r}}\ \delta\rho (\vec{r},t),
\end{equation}
where the particle density variation $\delta\rho (r,t)$ is
due to the external field $U_{ext}(t)$ of Eq. (\ref{eq1}).

A small isovector variation of the distribution function $\delta f \equiv
\delta f_{n} - \delta f_{p} $ can be evaluated using the linearized
collisional Landau-Vlasov equation with a collision term treated within the
relaxation time approximation \cite{dikola99}
\begin{equation}
\label{eq3}
\frac{\partial}{\partial t}\delta f +\vec{v}\cdot\vec{\nabla}_{\vec{r}} \delta f -
\vec{\nabla}_{\vec{r}} \left( {\delta U_{self} + \delta U_{ext}} \right) \cdot \vec{\nabla}_{\vec{p}} f_{eq} 
= - \left.\frac{1}{\tau_1}\delta f\right|_{l=1} - \left.\frac{1}{\tau_2}\delta f\right|_{l\ge 2},
\end{equation}
where $\vec{v}=\vec{p}/m^{\ast}$ is the velocity, $\tau_{k} $ is the relaxation 
time and $\delta U_{ext} \equiv U_{ext,n} - U_{ext,p}$. The notations $l=1$ and 
$l\ge 2$ mean that the perturbation of $\delta f|_{l=1}$ and $\delta f|_{l\geq 2}$ 
in the collision integral includes only Fermi surface distortions with a multipolarity 
$l=1$ and $l\ge 2$ in order to conserve the particle number in the collision 
processes \cite{abkh59}.

The variation of the isovector selfconsistent mean field $\delta U_{self}$
can be expressed in terms of the Landau's interaction amplitude
$v_{int}(\vec{p},\vec{p}^{\prime})$ as \cite{abkh59}
\begin{equation}\label{self1}
\delta U_{self}=\frac{1}{N(T)}\int \,\frac{g\,d\vec{p}^{\prime}}
{(2\pi\hbar)^{3}}\,v_{int}(\vec{p},\vec{p}^{\prime})\,\delta
f(\vec{r},\vec{p}^{\prime};t),
\end{equation}
where $N(T)$ is thermally averaged density of states. The
interaction amplitude $v_{int} \left(p, p^{\prime} \right)$ is
parameterized in terms of the Landau constants $F_l^{\prime} $ as
\begin{equation}
v_{int}(\vec{p},\vec{p}^{\prime})=
\sum_{l=0}^{\infty}\,F_{l}^{\prime}\,P_{l}(\hat{p}\cdot \hat{p}^{\prime}),
\qquad \hat{p}=\vec{p}/p.
\label{int1}
\end{equation}

The solution of Eq. (\ref{eq3}) can be found in the form of a plane wave
\begin{equation}
\delta f(\vec{r},\vec{p};t)=-{\frac{\partial f_{\mathrm{eq}}}
{\partial\epsilon_{p}}}\nu_{\omega,\vec{q}}(\vec{p})\,e^{i(\vec{q}\cdot
\vec{r}-\omega t)},
\label{deltaf1}
\end{equation}
where $\left.\partial f_{eq}\right/\partial\varepsilon_p$ is a sharply peaked
function at $p=p_F$. In case of the longitudinal excitation modes, the
Fermi-surface distortion function $\nu_{\omega ,q} (p)$  depends only on the 
angle $\theta$ between $p$ and $q$ and can be expanded in Legendre 
polynomials as
\begin{equation}
\nu_{\omega ,\vec{q}}(\vec{p})=\sum_{l=0}^{\infty}P_{l} (\cos \theta_{pq})\,
\nu_{l}(p),  
\label{deltaf2}
\end{equation}

Performing transformations in the same manner as in \cite{kolukh07}, one can 
come to the following set of equations for the amplitudes $\nu_l(p)$:
\begin{eqnarray}
\nu_{l}(p)+(2l+1)\sum\limits_{l^{\prime}=0}^{\infty}\frac{F_{l^{\prime}}^{\prime}}
{2l^{\prime}+1}\widetilde{\nu}_{l^{\prime}}Q_{ll^{\prime}}(z)
- \lambda_{0}(2l+1)Q_{l0}(z) \nonumber \\
=i(2l+1)\gamma \nu _{0}(p)\frac{1}{z}\left[ \delta _{l0}-Q_{l0}(z)\right]
-i(2l+1)\eta \gamma \nu _{1}(p)Q_{l0}(z).
%\label{eq3}
\end{eqnarray}

Here, $\eta=1-\left.\tau_2\right/\tau_1$, $\widetilde{\nu}_l$ is the averaged amplitude
\begin{equation}\label{eq9}
\widetilde{\nu}_l= -\frac{1}{N(T)}\int \frac{gdp}{(2\pi\hbar)^3}\, 
\frac{\partial f_{eq} ( \varepsilon_p)}{\partial \varepsilon_p} \nu_l (p),
\end{equation}
and
$$
Q_{ll^{\prime}}(z)=-\frac{1}{2}\int\limits_{-1}^{1}dx\frac{P_l(x)\ x\ P_{l^\prime}(x)}{z-x},
\quad x=\cos\theta_{pq},
\quad \gamma=\frac{1}{\tau_{2}qv},
\quad z=s+i\gamma,
\quad s=\frac{\omega}{qv}.
$$

For simplicity we will assume: $F_{l = 0}^{\prime}\ne 0$, $F_{l =1}^{\prime}\ne 0$, 
$F_{\ell \ge 2}^{\prime}  = 0$. Finally we obtain the density-density response function 
$\chi (\omega ,q)$ for a given momentum transfer $q$ in the following form
\begin{equation}
\label{eq10}
\chi \left( {\omega ,q} \right) = 2N(T)\frac{{\mathop
{\tilde {\chi} }\nolimits_{in} \left( {\omega ,q} \right)}}{{1 +
F_{0}^{\prime}  \mathop {\tilde {\chi} }\nolimits_{in} \left( {\omega ,q}
\right)}},
\end{equation}
where $\mathop {\tilde {\chi} }\nolimits_{in} \left( {\omega ,q} \right)$ is
the internal response function which depends on the temperature and the
relaxation time. The explicit form of
 $\mathop {\tilde {\chi}}\nolimits_{in} \left( {\omega ,q} \right)$ for 
$T=0$ and no relaxation is given in Ref. \cite{kolukh07}.

For finite nuclei, the boundary condition can be taken as a condition for
the balance of the forces on the free nuclear surface: 
$\vec{n}\cdot\vec{F}\vert_S+\vec{n}\cdot\vec{F}_S=0$, where 
$n$ is the unit vector in the normal direction to the nuclear surface $S$, the 
internal force $F$ is associated with the isovector sound wave and $F_S$ 
is the isovector surface tension force. Both forces 
$\vec{n}\cdot\vec{F}\vert_S$ and $\vec{n}\cdot\vec{F}_S$ can 
be represented in terms of isovector shift of the nuclear surface and the boundary 
condition takes the final form of the following secular equation \cite{kolukh07}
\begin{equation}
\label{eq11}
\left[ -\frac{1}{2}C_{sym} \mathop {\overline{\rho} } \nolimits_{eq}
- \frac{2}{3}\mu_{F} + \frac{2}{x^2} \mu_F  \right] j_1 (x) 
+ \left[-\frac{2}{x} \mu_F + \frac{4}{3} \frac{\rho_{eq}}{qr_{0} \left( 1 + \kappa_{NM}\right)} 
Q_{sym}  \right] j_1^{\prime} (x) = 0.
\end{equation}
Here $\mathop {\overline {\rho} } \nolimits_{eq} = \left( {\rho _{eq,n} +
\rho _{eq,p}}  \right)/2$, $x = qR$, $Q_{sym}$ is the effective isovector surface 
stiffness \cite{mysw74}, and
$$
\mu_F = \frac{3}{2} \varepsilon_F \rho_{eq} \frac{s_R^2}{1+\left.F_1^{\prime}\right/3}
\left[1-\frac{\left(1+F_0^\prime \right)\left(1 + F_{1}^{\prime}/3\right)}
{3\,s_R^2 }\right],
\quad
s_R =\frac{\omega_R}{v_F q}.
$$

\section{ Energy-weighted sums and transport coefficients}

The presence of the nonlocal interaction in Eq. (\ref{eq10}) gives rise to some
important consequences for the properties of the energy-weighted sums (EWS)
$m_k(q)$ for isovector mode. Let us introduce the strength
function per unit volume $S(\omega ,q)=\left.\textrm{Im}\chi (\omega ,q)\right/\pi$. 
The energy weighted sums are defined by
\begin{equation}
m_k(q)=\int\limits_0^\infty d(\hbar \omega )\ (\hbar \omega )^k\ S(\omega ,q).  
\label{mk1}
\end{equation}

In the case of cold nucleus $T = 0$ and no relaxation $\tau_1 , \tau_2\to\infty $,
 we recover well-known results \cite{list89}
\begin{equation}
\label{eq13}
m_{ - 1}^{\left( {0} \right)} \left( {q} \right) = \frac{{\rho _{eq}
}}{{2C_{sym}} },\quad m_{1}^{\left( {0} \right)} \left( {q} \right) = \hbar
^{2}\frac{{\rho _{eq}} }{{2m^{\prime} }}q^{2},\quad m_{3}^{\left( {0}
\right)} \left( {q} \right) = \hbar ^{4}\frac{{C_{sym}^{\prime}  \rho _{eq}
}}{{2m^{\prime 2}}}q^{4},
\end{equation}
where $\rho _{eq} $ is the equilibrium particle density, $C_{sym}$ is the 
isospin symmetry energy, $\varepsilon _{F} $ is the Fermi energy and $m^{\prime} =m^{*} 
\mathord{\left/ {\vphantom {{m^{ *} } {\left( {1 + F_{1}^{\prime}  /3}
\right)}}} \right. \kern-\nulldelimiterspace} {\left( {1 + F_{1}^{\prime}
/3} \right)}$ is the effective mass for isovector mode, $m^{ *}  = m\left(
{1 + F_{1} /3} \right)$. The renormalized symmetry energy $C_{sym}^{\prime}$ 
in Eq. (\ref{eq13}) is given by $C_{sym}^{\prime}=C_{sym}+8\ \epsilon_{F}/15$ 
where the last term is due to the Fermi surface distortion effect \cite{kosh01}.

In contrast to the isoscalar mode, the isovector EWS $m_1^{(0)} (q)$ of Eq. 
(\ref{eq13}) is model dependent. As can be seen from Eq. (\ref{eq13}), the sum 
$m_1^{(0)} (q)$ includes the enhancement factor for infinite nuclear matter 
$1+\kappa_{NM} = (m/m^{ *}) (1+\left.F_1^{\prime}\right /3)$,
which depends on the nonlocal interaction constant $F_1^{\prime}  \ne 0$.

\section{Results and Discussions}

In this work we have adopted the value of $r_0=1.2\ fm$ and the effective nucleon mass 
$m^{ *} $ was taken as $m^{ *}  = 0.9m$\textbf{} which corresponds to the Landau parameter
$F_{1} = - 0.3$. For the isovector interaction parameter $F_{0}^{\prime}$ we have used 
$F_{0}^{\prime }=1.41$ to keep a reasonable value of isospin symmetry energy $C_{sym} $ 
of the order of 60 MeV. The interaction parameter $F_{1}^{\prime}  $ can be estimated by
considering the enhancement factor in the isovector EWS.

The relaxation time $\tau _{k} $ in Eq. (\ref{eq3}) is frequency and temperature
dependent. We will assume the following form of $\tau _{k} = {{\hbar \alpha
_{k}}  \mathord{\left/ {\vphantom {{\hbar \alpha _{k}}  {\left[ {T^{2} +
\left( {\hbar \omega /2\pi}  \right)^{2}} \right]}}} \right.
\kern-\nulldelimiterspace} {\left[ {T^{2} + \left( {\hbar \omega /2\pi}
\right)^{2}} \right]}}$ [5]. The parameter $\alpha_k$ depends on the $NN$-scattering 
cross sections. We will adopt $\alpha _{2} = 5.4$ MeV, which corresponds to the in-medium
$NN$-scattering cross sections.

Following Ref. \cite{dikola99}, we derive the photoabsorption cross section 
$\sigma_{abs}(\omega)$ in terms of the strength function $S(\omega ,q)$ as follows
\begin{equation}
\label{eq14}
\sigma _{abs} \left( {\omega}  \right) = \frac{{4\pi ^{2}e^{2}}}{{cq^{2}\rho
_{eq}} }\frac{{NZ}}{{A}}\omega S\left( {\omega ,q} \right)\;.
\end{equation}

In the case of the velocity independent $NN$-interaction, the cross section
$\sigma_{abs}(\omega)$ is normalized by the ordinary Thomas-Reiche-Kuhn 
sum rule \cite{risch80} (see $m_1^{(0)}(q)$ in Eq. (\ref{eq13}) for $\kappa_{NM}=0$)
\begin{eqnarray}
\widetilde{m}_{1,\mathrm{TRK}}^{(0)}=\int\limits_0^\infty\ d(\hbar\omega)\
\sigma_{\mathrm{abs}}(\omega) 
\nonumber \\
=\frac{2\pi ^{2}\hbar e^{2}}{mc}
\frac{NZ}{A}
\quad \textrm{for}
\quad T=0,\  \kappa_{NM}=0,
\quad \textrm{and}
\quad \tau_{1}, \tau _{2}\rightarrow\infty.
\label{mk5}
\end{eqnarray}

Taking into account the velocity dependence of the $NN$-interaction with $F_1 \ne 0$ 
and $F_1^\prime \ne 0$, we note that both the enhancement factor $\kappa_{NM}\ne 0$ 
in $m_1^{(0)} (q)$ of Eq. (\ref{eq13}) and the corresponding correction at the last term 
of the boundary condition affect the sum rule (\ref{mk5}). For $\kappa_{NM}\neq 0$, we 
obtain the following result \cite{kolukh07}
\begin{eqnarray}
\widetilde{m}_1^{(0)}=\int\limits_0^\infty\ d(\hbar\omega)\
\sigma_{\mathrm{abs}}(\omega) \nonumber \\
=\frac{2\pi^2\hbar e^2}{mc}
\left(\frac{q_1^\prime(A)}{q_0(A)}\right)^2 \frac{NZ}{A}(1+\kappa_{NM})
\quad \textrm{for}
\quad T=0,\  \kappa_{NM}>0,
\quad \textrm{and}
\quad \tau_{1}, \tau _{2}\rightarrow\infty,
\label{mk6}
\end{eqnarray}
where $q_0(A)$ and $q_{1}^{\prime}(A) $ are the lowest roots of the boundary condition equation
(\ref{eq11}) for $\kappa _{NM} = 0$ and $\kappa _{NM} \ne 0$, respectively.

\begin{figure}
\begin{center}
\includegraphics*[width=12cm]{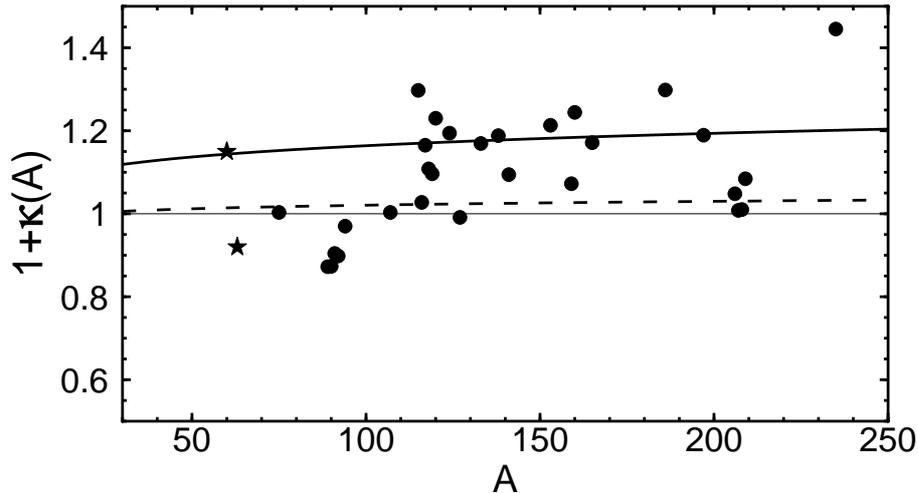}
\end{center}
\caption{Dependence of the enhancement factor $1+\protect\kappa (A)$ for IVGDR
on the mass number $A$. The result of calculation was obtained for the Landau's
amplitude $F_{1}=-0.3$, $F_{1}^{\prime}=1.1$ (solid curve) and $F_{1}=-0.3$,
$F_{1}^{\prime}=0$ (dashed curve). The points are the experimental data of the
Livermore group from \cite{befu}. Two points noted by symbol "$\star$" were
obtained by the inclusion of the contribution from the ($\gamma,p$) cross section
\cite{be}.}
\label{fig1}
\end{figure}

\figurename\ \ref{fig1} shows the dependence of the enhancement factor 
$\left.\widetilde{m}_1^{(0)} \right/ \widetilde{m}_{1,TRK}^{(0)}=1+\kappa(A)$ 
on the mass number $A$ for two interaction parameters $F_1^{\prime} $ in comparison 
to the experimental data from Refs. \cite{befu,be}. We see that the enhancement factor for 
the IVGDR sum rule is $A$-dependent. This $A$-dependence is due to the boundary 
condition. Our estimate of the enhancement factor $\kappa(A)$ is about 10\%
for light nuclei and increases to 20\% for heavy nuclei. We point out that
about of 5\% of the EWS enhancement is caused by the dependence of the effective
nucleon-nucleon interaction on the nucleon velocity in the isoscalar channel
(dashed curve in \figurename\ \ref{fig1}). The value of interaction parameters 
$F_1^{\prime} $ can be obtained from a fit of the evaluated enhancement factor 
$\kappa(A)$ to the experimental data. In this work we have used
$F_1^\prime =1.1$.

\begin{figure}
\begin{center}
\includegraphics*[width=10cm]{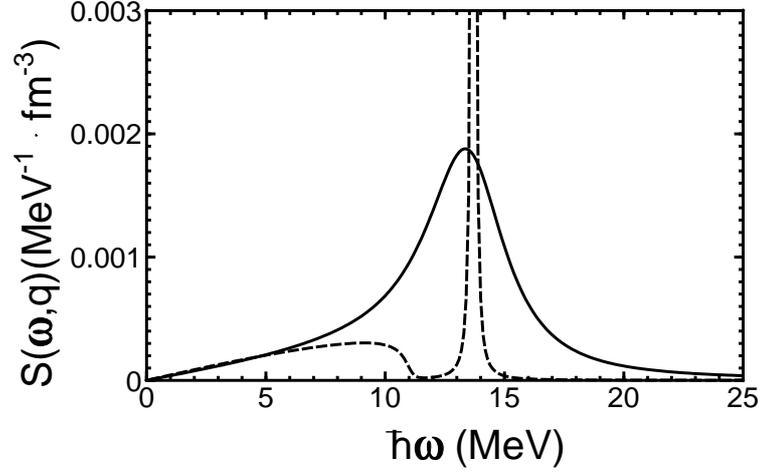}
\end{center}
\caption{Strength function $S(\omega,q)$ for $F_{1}=-0.3$, $F_{1}^{\prime}=1.1$, 
$\eta=2/3$, $A=208$. Solid line for $T=5$ MeV, $\alpha_{2}=5.4$ MeV and dashed 
line for $T=0.5$ MeV, $\alpha_{2}=100$ MeV.}
\label{fig2}
\end{figure}

Performing the numerical calculations of the response function $\chi (\omega ,q)$ 
(\ref{eq10}), one can evaluate the strength function $S(\omega ,q)$ and the 
energy-weighted sums $m_k(q)$
for $T \ne 0$ and in presence of relaxation. The strength function
$S(\omega,q)$ is sensitive to the
interaction parameters and to the relaxation properties. Because of
$F_{0}^{\prime}  > 0$, the IVGDR strength function contains both the sound
mode contribution at $s > 1$ and the Landau damping region at $s < 1$. This
is illustrated in \figurename\ \ref{fig2}.

The presence of the Landau damping in the IVGDR $S (\omega ,q)$ is well seen in 
\figurename\ \ref{fig2} for the zero-sound regime $\omega_R\tau_2 >>1$ (dashed 
line) as a wide bump on the left side of the narrow sound peak. For high temperature 
(solid line in \figurename\ \ref{fig2}), the sound peak becomes wider due to the decrease 
of the relaxation time (collisional relaxation), and due to the collisionless thermal Landau 
damping which increases with $T$. As can be seen from \figurename\ \ref{fig2}, 
overlapping of both the sound peak and the Landau damping bump leads to the 
asymmetry of the IVGDR resonance at high temperatures.

\begin{figure}
\begin{center}
\includegraphics*[width=10cm]{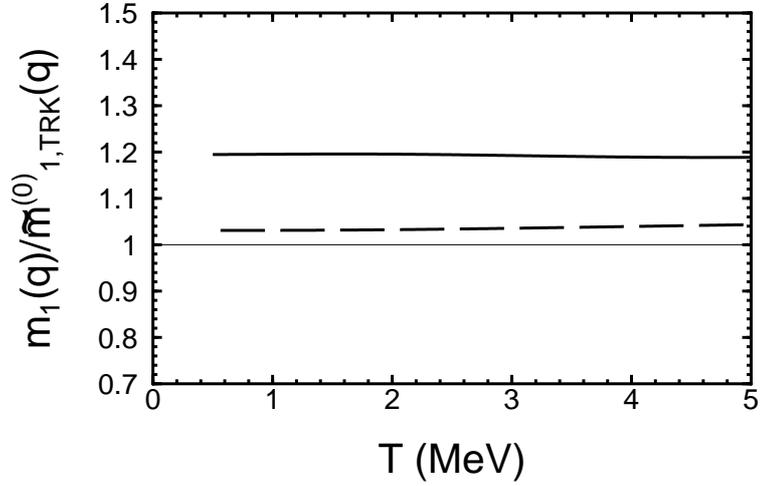}
\end{center}
\caption{Temperature dependence of the "model independent" EWS $m_{1}(q)$
normalized to the value $\widetilde{m}^{(0)}_{1,\mathrm{TRK}}$ from Eq. (15). 
The calculations performed for the nucleus $A=208$ using $\alpha_2=5.4$ MeV, 
$\eta=1$ and $F_1=-0.3$. Solid line for $F_1^{\prime}=1.1$ and dashed line 
for $F_1^{\prime}=0$.}
\label{fig3}
\end{figure}

We have studied the temperature behaviour of the "model independent" EWS
$m_1 (q)$ and the enhancement factor 
$\left.m_1 (q)\right/\widetilde{m}_{1,\mathrm{TRK}}^{(0)}$. 
For non-zero temperatures and in presence of the relaxation, the energy-weighted
sum $m_1 (q)$ has been evaluated using the definition (\ref{mk1}) and the response 
function $\chi (\omega ,q)$ from Eq. (\ref{eq10}). In \figurename\ \ref{fig3} we have 
plotted the ratio $\left.m_1(q)\right/\widetilde{m}^{(0)}_{1,\mathrm{TRK}}(q)$ 
as a function of temperature $T$. We can see from \figurename\ \ref{fig3} that the 
enhancement factor is only slightly sensitive to the temperature variation.

\section{Summary and Conclusions}

Starting from the collisional kinetic equation (\ref{eq3}), we have derived the
strength function and the energy-weighted sums $m_k$ for the isovector 
excitations in the heated nuclear matter and the finite nuclei. An important 
ingredient of our consideration is the inclusion of the velocity dependent 
$NN$-interaction for both the isovector and the isoscalar channel simultaneously 
providing the isoscalar effective mass $\left.m^{ *}\right/m <1$ and the enhancement 
factor $1+\kappa(A) >1$ for the isovector "model independent" sum $m_1$. Our
 consideration is valid for arbitrary collision parameter $\alpha_k$ and can 
be used, particularly, for the transition region from the zero sound (collisional)
regime to the first sound (hydrodynamic) regime.

We have adopted a simple Fermi liquid drop model with two essential
features: 
\begin{enumerate}
\item[(i)] The linearized kinetic equation is applied to the
nuclear interior, where the relatively small oscillations of the particle
density take place;
\item[(ii)] The dynamics in the surface layer of the nucleus is
described by means of the macroscopic boundary condition which is taken as a
condition for the balance of the forces on the free nuclear surface.
\end{enumerate}
This model provides a satisfactory description of the $A$-dependence of the 
enhancement factor of the IVGDR sum rule, see \figurename\ \ref{fig1}.
 Moreover, the value of interaction
parameters $F_1^\prime \approx 1.1$ was derived from a fit of the
evaluated enhancement factor in \figurename\ \ref{fig1} to the experimental data.

The main goal of present paper is the calculations of the strength function
$S (\omega ,q)$ and the energy-weighted sums $m_k(q)$ for finite temperatures 
$T>0$ and in presence of relaxation processes. We have shown that the Landau 
damping effect occurs in the isovector $S(\omega ,q)$ at low temperatures as a 
wide bump on the left side of the narrow sound peak (see \figurename\ \ref{fig2}). 
For high temperature, the overlapping of both the sound peak and the Landau 
damping bump leads to the asymmetry of the IVGDR resonance. The isovector 
EWS shows only minor temperature dependence. In particular, the "model independent" 
EWS $m_1 (q)$ and the corresponding enhancement factor 
$\left.m_1(q)\right/\widetilde{m}^{(0)}_{1,\mathrm{TRK}}(q)$ 
are practically constant in the interval of temperature $T = 0\, \div \,5$ MeV,
see \figurename\ \ref{fig3}.

\end{document}